# COVID-19 Datathon Based on Deidentified Governmental Data as an Approach for Solving Policy Challenges, Increasing Trust, and Building a Community: Case Study

Mor Peleg, PhD[1], Amnon Reichman, SJD[2], Sivan Shachar, Adv[2], Tamir Gadot, MA[1], Meytal Avgil Tsadok, PhD[3], Maya Azaria, Adv[4], Orr Dunkelman, PhD[1,2], Shiri Hassid, PhD[3], Daniella Partem, MA[4], Maya Shmailov, PhD[5], Elad Yom-Tov, PhD[6], Roy Cohen, MEng[3]

[1]Data Science Research Center, The University of Haifa, Haifa, Israel

[2]The Center for Cyber Law & Policy, The University of Haifa, Haifa, Israel

[3]Timna - Big Data Platform unit, Ministry of Health, Israel

[4]The Center for the 4th Industrial Revolution, Israel innovation Authority, Israel

[5]Shenkar College of Engineering and Design

[6]Microsoft Research, Israel

Corresponding author:

Prof. Mor Peleg
Editor in Chief, Journal of Biomedical Informatics
Director, University of Haifa's Data Science Research Center
Dept. of Information Systems
University of Haifa, 3498838, Israel
Rabin Building, room 7047
Email: morpeleg@is.haifa.ac.il
Web-site: is.haifa.ac.il/~morpeleg

**Abstract**.

Triggered by the COVID-19 crisis, Israel's Ministry of Health (MoH) held a virtual Datathon based on deidentified governmental data. Organized by a multidisciplinary committee, Israel's research community was invited to offer insights to COVID-19 policy challenges.

The Datathon was designed to (1) develop operationalizable data-driven models to address COVID-19 health-policy challenges and (2) build a community of researchers from academia, industry, and government and rebuild their trust in the government. This study also outlines a process for future data-driven regulatory responses for health crises.

Three specific challenges were defined based on their relevance (significance, data availability, and potential to anonymize the data): immunization policies, special needs of the young population, and populations whose rate of compliance with COVID-19 testing is low. The MoH team extracted diverse, reliable, up-to-date, and deidentified governmental datasets for each challenge. Secure remote-access research environments were established using Amazon Web Services, and data science tools were installed, and data security was tested. Registration was open to all citizens. The MoH screened the applicants and accepted around 80 participants, teaming them to balance areas of expertise as well as represent all sectors of the community. One week following the event, anonymous surveys for participants and mentors were distributed to assess overall usefulness and points for improvement and retention for future datathons.

at the 48-hour Datathon and pre-event sessions included 18 multidisciplinary teams, mentored by 20 data scientists, 6 epidemiologists, 5 presentation mentors, and 12 judges. The insights developed by the 3 winning teams are currently considered by the MoH as potential data science methods relevant for national policies. The most positive results were increased trust in the MoH and greater readiness to work with the government on these or future projects. Detailed feedback offered concrete lessons for improving the structure and organization of future government-led datathons.

Keywords: COVID-19; Hackathon; Datathon; Evidence-based regulation; Privacy; Data; Health policy; Trust; Public confidence; Public engagement, Public-private interface

# 1. Introduction

The COVID-19 pandemic has shaken the daily lives of many. Societies around the globe have been facing unprecedented challenges to which governments have responded with myriad national and local health care policies and regulations that ultimately rest on citizens' compliance. Since such policies entail significant infringements on basic liberties, the elementary requirement to base the regulatory schemes on sound evidence is a prerequisite for public confidence and trust, and of social resilience. The push towards evidence-based regulation, especially in the face of uncertainty, pressed state agencies to mass collect data, intensifying concerns related to privacy, dignity, and equity. Transparency and accountability, ever more important, had to be balanced against the commitment to protect private information. Against this complex backdrop, COVID-19 data policies play a central role.

The Ministry of Health of Israel (MoH) has been providing up-to-date national data on the COVID-19 pandemic via DataGov (Israeli government data set archive) [1] and Corona Dashboard [2], allowing the research community in the health care sector, academia, and Israeli industry to conduct research in the field of health. Due to the structure of Israel's health care system, the government has a significant amount of the relevant data at their disposal. These unique platforms enable the analysis and cross-referencing of medical, demographic, and other data, while complying with privacy regulations.

In early March 2021, a year after COVID-19 regulations started to be enforced locally, Israel has been fighting its third wave of outbreak by operating its large-scale immunization plan, limiting citizens' ability to gather, and restricting lower and high schools to open only at times and locales of low COVID-19 incidence, while grades 7-10 remained studying from home via Zoom (Figure 1).

To tackle the rapidly moving COVID-19 pandemic, TIMNA, the Big Data Platform unit at the MoH, initiated a COVID-19 "Datathon" [3,4]—a hackathon that revolves around data and utilizes data science methods. The Datathon's goal was to provide prototype models ("insights") for pressing COVID-19-related policy challenges by engaging the research community from all sectors: high-tech industry, health care, academia, and the public sector. To do so, the MoH partnered with the Innovation Authority (InnovA), research centers at the University of Haifa, and industry (primarily Microsoft), thereby forming a diverse organizing team. The organizing team, who are the coauthors of this paper, helped the MoH identify specific contemporary challenges which could be solved using data, and assisted in operationalizing the Datathon (See Section 2.4). These related to immunization and testing strategies as well as to opening of schools.

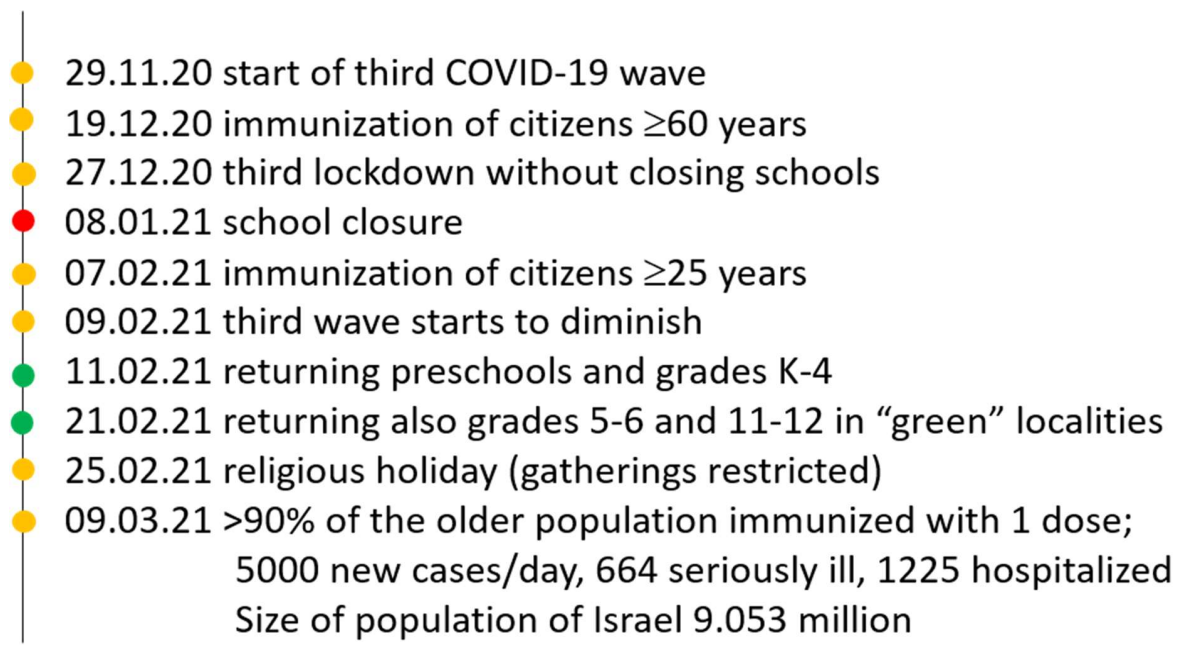

Figure 1. Important regulatory events in handling COVID-19 pandemic in Israel close to the start of the Datathon

The Datathon had three objectives:

1. To assist the government in shaping effective instruments and policies for addressing COVID-19 challenges by developing timely, relevant, and implementable data-driven insights.

2. To promote a trust-based, fertile, collaborative community of expert researchers from multidisciplinary sectors, who could team in an accountable and transparent process to brainstorm and rapidly develop data-driven insights to address national and international health care challenges.

3. To develop a process for future similar events, based on deidentified governmental data, that are feasible, useful, and compatible with rule-of-law and civil rights guarantees.

The datathon format was chosen to achieve these objectives because it is premised on sharing live data in a controlled, secured environment and provides teams with peer review from mentors and evaluation from judges and with the opportunity for exposure to decision-makers. This could result in subsequent opportunities to work together with the MoH by applying the suggested insights in practice. While many other COVID-19-related hackathons had already taken place [4-12], this Datathon is the first that was initiated by the government, engaged all sectors of society, and allowed them to use up-to-date governmental data.

This paper aims to share the methods that we designed for organizing a virtual Datathon. We hope that other governments and agencies will find these methods useful and that they can learn from our experience and engage the participation of multiple stakeholders, thereby benefiting from diversity of approaches to challenges based on wide-ranging, reliable, and deidentified governmental databases. Such an approach holds the potential of increasing the public's trust in the government and its regulatory policies.

## 2. Methods

To meet the objectives, an accountable and transparent process was used to plan and carry out the Datathon (Table 1 and Figure 2). Validation of the steps taken was done via an anonymous questionnaire distributed to participants and mentors one week following the event.

### 2.1 Initiating the idea and early steps

TIMNA, a representative from Microsoft and a design scholar from Shenkar College, conducted extensive consultations with relevant health care–field stakeholders. Pursuant to those discussions, the MoH conceived of the idea to conduct a Datathon six months prior to the event. A team of 6 professionals was established, and they conducted 11 interviews with national and international data consumers, providers, and users during August to September 2020. Among the interviewees were health organizations, hospitals, the nursing division at MoH, municipalities, and welfare services. The interviews focused on three main themes: (1) privacy hurdles, (2) lack of trust with other agencies and industry, and (3) poor communication and ability to easily access the data.

Two months prior to the event, a diverse organizing committee was formed from the MoH, academia (University of Haifa (UoH): Center for Cyber Law & Policy (CCLP) and Data Science Research Center (DSRC)), the high-tech industry (Microsoft), and the National Innovation Authority (InnovA). The event itself took place virtually, during March 9-11, 2021. Registration was open to the Israeli public and was free of charge.

Table 1. Steps in the Datathon organization (and responsible partners). Recommended new steps and changes based on lessons learned are shown in blue.

| Time | Task |
| --- | --- |
| (6mo-2mo) pre-event | Project inception by government (MoH) with multiple stakeholders.<br><br>Recruitment of Organizing Committee.<br><br>MoH team thought of 6 general topics; interviewed stakeholders from industry, academia, and government; and ranked topic importance considering (a) significance, (b) data availability, and (c) ability to deidentify data that would allow interesting analyses to be conducted. |
| (9wk-8wk) pre-event | Preparation of privacy-preserving standardized data sets (MoH with advice from UoH).<br><br>Preparation of Datathon website and publicity via MoH spokesperson, social media, and mailing lists related to AI in Healthcare (MoH, InnovA).<br><br>Preparation of registration Google Forms for participants and mentors (UoH). *Participant registration forms* included name, email, cellular phone #, professional link (e.g., LinkedIn), company/institution, position/field of studies, professional expertise (health care practitioner, health manager, product developer, business entrepreneur, engineer, statistician, data scientist, law/policy professional, other + indication of the number of years of experience), whether the registrant is part of a team, which challenge the registrant is interested in, suggestions for data needed for the challenge, whether any organization has ever disqualified their access to any database or data? If so, what were the circumstances?, and commitment to continue into an acceleration process with the Datathon product/idea.<br><br>*Mentor registration forms* included name, email, cellular phone #, professional link (e.g., LinkedIn), company/institution, position/field of studies, head photo, mentoring expertise (all that apply from data science, clinical/epidemiology, law, and presentation), availability during the scheduled 4 mentoring sessions (check all that apply), and additional comments.<br><br>List of mentors and judges (data science, law, epidemiology, and presentation) led by UoH and approved by MoH.<br><br>Preparation of the Terms and Conditions for participation (UoH). |
| (at least 6wk recommended)<br>(5wk-4wk) pre-event | Publicizing the event and opening registration for at least 3 weeks recommended (2.5 weeks).<br><br>New emphasis: disclose the limitations on the number of participants, and the lack of budget for prizes and food. |
| (4wk-3wk) pre-event<br>(4wk-2wk) pre-event | Preparation of the judging criteria and tables for projects rating (UoH); the criteria related to the project's importance and potential impact (15%), methodology (25%), innovation (20%), ease of implementation and compliance with regulatory regimes (15%), potential for pursuing further R&D with the team (5%), clarity of the presentation (5%), overall impression (15%).<br><br>Preparation of the detailed event schedule (UoH).<br><br>Scheduling mentors for mentoring sessions and judges for the challenge judging and final judging sessions (UoH). |
| (4wk-3wk) pre-event<br><br>(3wk-2wk) pre-event | Design of the tables and their presentation to stakeholders to make sure that the data would be valuable for analysis (MoH).<br><br>Updates and first iterations of data queries. Operationalization of first data deidentification scheme (MoH). |
| (3wk-2wk) pre-event<br><br>2wk pre-event | Selection process (MoH).<br><br>Matchmaking event for participants seeking teams (MoH).<br><br>Final formation of teams (MoH).<br><br>Creating of tutorials on the data and virtual machine environments (MoH).<br><br>Setup of communication routes: Slack channels and Zoom rooms (UoH).<br><br>New step: Pilot testing with one group per challenge to ensure the quality of the data set and the virtual research environments.<br><br>Preparation of a plan for publicity over public media. |
| 1wk pre-event | Three Zoom workshops with mentors, judges, participants (everyone) to explain the challenges and data sets along with the data dictionary of each dataset, with variable names and explanations; Q&A, address ethical considerations, and answer questions; the workshop with mentors should cover realistic expectations.<br><br>Testing the participants' ability to join the data environments and communication in Slack (everyone).<br><br>Q&A via Slack (MoH+UoH).<br><br>Additional iterations of data queries and operationalization of the complete deidentification scheme following data inspection (MoH).<br><br>Production of the final data sets using the formulated queries and the deidentification scheme and depositing them in the virtual research environment (MoH). |
| The Datathon itself | Opening and Closing events led by InnovA and MoH.<br><br>Technical support for virtual environments 24/7 (MoH, Amazon) with installation of requested software updates every 24 hours.<br><br>Monitoring for privacy issues (MoH, UoH).<br><br>General technical support (UoH).<br><br>Communication via Slack and Zoom (led by UoH).<br><br>Communication with mentors via Mentornity.<br><br>Status updates should include some suggestions from mentors that may be helpful for a given challenge; teams should be made aware ahead of time that they would be requested to present themselves during the Status Update meetings to support community building.<br><br>Moderators for judging sessions (UoH).<br><br>Communicating feedback from judges to the teams.<br><br>Sending T-shirts to participants (UoH). |
| 1wk after the event | Participants and mentors survey (led by UoH). |

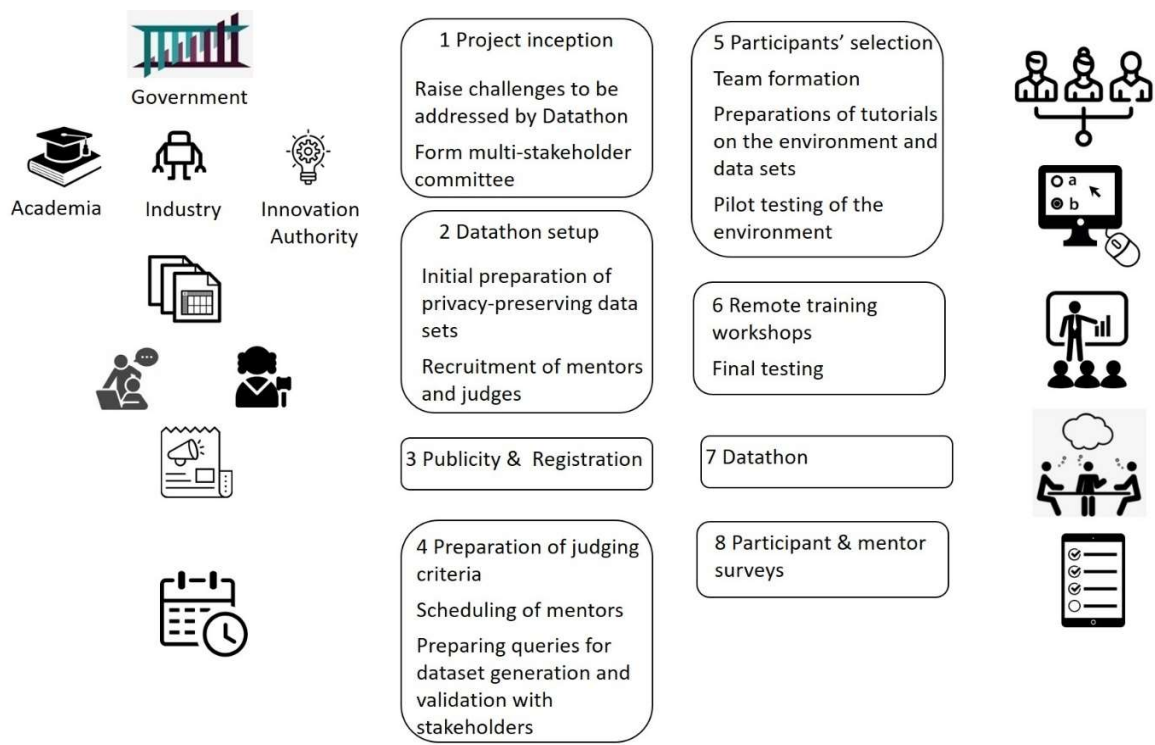

Figure 2. Process of Datathon organization

## 2.2 The regulatory framework

Attention was paid to the regulatory framework relevant to the Datathon. Data protection (and related privacy concerns), were addressed by the design of the environment and through the bylaws governing the activity, which were integrated into the registration process so that each participant had to undertake a legally binding commitment not to engage in reidentification or otherwise breach data and privacy laws.

The Datathon offered the MoH an opportunity to engage with data science talent to develop better policies for handling the COVID-19 pandemic. The judges' panel included four members of the operative team tasked with managing the pandemic ("The Corona Cabinet").

Since the Datathon focused on generating data-driven policy tools for governmental decision-makers, and since such tools were oriented to be considered as possible components of regulatory regimes, legal approval of ethical boards according to the Helsinki protocols was not formally required. Special attention was paid to the matter of intellectual property (IP), and the terms of participation in the Datathon provided that the MoH (and the Israeli government more generally) will not be required to pay for the use of any outcome of the Datathon, but IP rights remain with the participants for commercialization in markets outside Israel, and vis-à-vis health organizations other than the Israeli government.

## 2.3 The challenges

To identify challenges that would be relevant at the timing of the event (two months later), The MoH team set up interviews with multiple stakeholders: senior employees within the MoH, researchers from policy-oriented centers, and leaders from the industry with emphasis on companies specializing in data. The MoH team presented the suggested challenges and asked the interviewee to rank them on a scale of 1-5 in 3 dimensions: (1) the degree of interest, (2) the availability of data to support analyses related to the challenge, and (3) the potential for deidentifying the data while balancing risks and retaining meaningful features.

The three highest ranked challenges that were relevant at the time (January 4, 2021) were (1) the vaccines' effectiveness, (2) the impact of school-reopening policies on the morbidity burden, and (3) identifying populations that are underrepresented in coronavirus tests. The definition of the challenges was refined, as Increasing the effectiveness of immunization strategies; Management of challenges at the young population; and Improving receptiveness to undergo COVID-19 tests.

## 2.4 Preparation: management team and process

The organizing committee had twice-weekly and additional meetings using Microsoft Teams. Trello was used for task planning and follow-up. WhatsApp groups were formed and used for communication. The steps for organizing the Datathon are summarized in Table 1 and Figure 2.

## 2.5 Data preparation and secure rooms setup

The MoH routinely shares data with the public in a manner compatible with contextual privacy principles. These methods include (a) tables with few fields to prevent crossing the data with other data sources in order to reidentify individuals, (b) values that are grouped into ranges (e.g., age groups rather than date of birth), and (c) not presenting data of groups of items belonging to less than 15 individuals, in accordance with the policies of the Israeli Bureau of Statistics. These modes of abstractions or dilution still provide important information, but limit privacy-related risks.

For the Datathon, data was selected based on the challenges posed (Appendix A). The set of variables selected for each challenge was chosen by an epidemiologist who was familiar with the clinical needs and available data. A deidentification and anonymization protocol was then implemented for each dataset to ensure minimal potential for identifying individuals. The potential tables were then presented to different researchers, and changes were applied based on their comments.

The data included details that have not been shared with the public before (Appendix A): detailed data on an individual level, with more granular age groups, data about the sector (Jewish, Arab, ultra-Orthodox), statistical area (zip), and socioeconomic level, along with data on the size of each locality and statistical area. These characteristics may reflect social norms relevant to the challenges or may otherwise impact potential reactions to policies that may be generated by the Datathon, but at the same time, these characteristics are sensitive, even at the aggregate level, and hence they had not been previously released to the general public.

The data that were shared were current, reflecting the situation 96 hours prior to the beginning of the event. Each group had access to the relevant table (in CSV format), with no access to the tables of the other challenges. In addition, the shared virtual environment provided access to more publicly available datasets such as updated tables from the Israeli government data set archive [1], publicly available tables from the Israeli Bureau of Statistics that could enrich the main datasets by linking to the geographic statistical areas, and other publicly available tables from different governmental offices.

The research environment was virtual, participants worked from home and groups worked together via virtual discussion channels. We used a secure virtual machine research environment provided by Amazon Web Services (AWS). Penetration tests were conducted to ensure no data breach could be made into the environment, and to minimize any risk of data extraction from the environment. The virtual environment applied a simple user-friendly connection using multifactor authentication via the Duo Mobile application. The software available in the research environment included: R, R-Studio, Python, Anaconda, and Jupyter notebook; Power BI; and MS Office tools.

## 2.6 Participants, mentors, and judges recruitment and team building

Google Forms for registration of participants and mentors were prepared. The fields included in the participant and mentor registration forms are specified in Table 1 (row 2).

The website created for the event included a link to the participant registration forms [13]. The event was advertised on social media (LinkedIn, Facebook, Twitter), on the MoH website, and on professional email lists, including lists of nonprofit organizations focusing on advancing women and minorities in high-tech and data science, utilizing the InnovA communications. An announcement was also sent to representatives of all academic data science research centers, who were asked to forward to relevant researchers. Registration was open to all Israeli citizens, free of charge. The response was more than anticipated, such that less than 28% of the applicants could participate.

The data from the registration forms included demographic, sector, and experience information of each participant, and were automatically fed into a spreadsheet. This allowed participants' screening by the MoH and professional auditors from InnovA. Subsequently, the candidates with the best expertise in data science, epidemiology, and regulation and policy were selected. The number of candidates selected was planned for 5-6 groups per challenge and around 4-5 participants per group, yielding around 80 participants (78). Participants who registered as a team remained as such and were also assigned to the challenge that they had indicated as preferred. The teams who self-organized were from high-tech companies, government, public organizations, or Academia. When groups were small (2-3 participants), individual registrants were added to complement the team's expertise. Based on the information provided in the spreadsheet, care was taken to include representation from all religious and ethnic sectors, males and females, young and older participants, experienced and more junior participants (e.g., students), different occupation sectors, including academia, industry, health care, and law/policy/governance, and different disciplines (data science, law and governance, epidemiology).

After the schedule for the Datathon was created (Table 2) and the number of sessions with mentors (4) and participants (~80) became clear, the number of mentors (36) for the event was determined such that each mentor would be available in at least one session and that all sessions would be covered. This resulted in 20 data science mentors; 6 epidemiology mentors; 5 presentation mentors; and 5 policy/regulation mentors.

The roles of the mentors were to (a) help teams with issues that arose during their research, (b) provide the teams with a broader view and help them brainstorm and think creatively, and (c) help teams maximize value from the data available to them during the 48 hours allocated.

Table 2. Datathon schedule

| **Day 1** | **Day 2** | **Day 3** |
|---|---|---|
| 13:00-14:00 Opening (Zoom) | 08:00-08:30 Nia workout (Zoom)<br><br>09:00-10:00 Status update (Zoom)<br><br>10:00-11:00 Presentation preparation workshop (Zoom) | 08:00-08:30 Nia workout (Zoom)<br><br>09:00-09:30 Status update (Zoom) |
| 14:00-15:30 Teams work without mentors | 11:00-13:00 Mentoring session (6 data science, 2 epidemiology, 1 law) Zoom | 09:30-13:00 Mentoring session (2 data science, 1 epidemiology, 1 law, 3 presentation) Zoom<br><br>Teams work and could consult with mentors |
| 15:30-16:30 Introductions to the data sets and Q&A (20 minutes per challenge, Zoom) | 15:00-18:00 Mentoring session (6 data science, 2 epidemiology, 1 law, 2 presentation*) Zoom<br><br>*presentation mentors were available from 16:00-19:00 | 13:00 Presentation submission<br><br>13:30-15:30 Semi-finals judging (1 room per challenge)<br><br>15:30-16:00 Judges select 2 finalists per challenge |
| 16:30-19:00 Mentoring session (6 data science, 2 epidemiology, 1 law) Zoom<br><br>Teams work and could consult with mentors | Teams work and could consult with mentors | 16:00-18:30 Finals and closing event |
| 19:30-20:30 Status update (Zoom) | 19:30-20:30 Status update (Zoom) | |

## 2.7 Projects evaluation by judges

The project-evaluation criteria were proposed by the UoH and were discussed and refined with the team of judges. The criteria related to the project's importance and potential impact (15%), methodology (25%), innovation (20%), ease of implementation and compliance with regulatory regimes (15%), potential for pursuing further R&D with the team (5%), clarity of the presentation (5%), and overall impression (15%). Upon receiving comments from the judges, these criteria were broken down into more concrete questions (Appendix B), in order to ensure common language between judges.

Participants were asked to create presentations using MS-Office tools available in the virtual environment and place them into a dedicated folder in that environment. Retrieval of the presentations was done 30 minutes before judging.

The evaluation was carried out in two steps (semi-finals and finals). During the semi-finals, which took 2 hours and was held on the last day, 48 hours after the start of the event, 3 Zoom rooms were available, one per challenge. Only the groups belonging to each challenge, 3 judges, a moderator (per challenge), and members of the organizing committee attended the semi-finals. Each team (6 teams were planned per challenge) presented for 7.5 minutes and answered the judges' questions from another 7.5 minutes. Each judge was given access to a secure and private Google Sheet, prepared according to the evaluation criteria. The judges were asked to rate the projects according to each criterion on a Likert scale of 1 through 5 and also include a verbal assessment. The assessments of each judge were automatically imported into a Google Sheet that was shared with the judges by a moderator from UoH, who projected the Google Sheet to all judges. The judges then selected 2 finalists, and the moderator passed the project names to the moderator of the finals. The judges had 45 minutes to arrive at their decision. The closing ceremony started 15 minutes before the judges completed their decision.

Six judges were scheduled to evaluate the 6 finalists. This time, all participants and invited audience members assembled into a single Zoom room. A procedure similar to that of the semi-finals was used for evaluation by the judges. The closing event continued during the 30 minutes of judging until the 3 winners were announced.

## 2.8 Tools for virtual communication

The Datathon was run at a time when public gatherings were restricted. It was therefore entirely virtual. To allow video conferencing, Zoom rooms were provided: a general room and three additional rooms, one per challenge—were created for the introductions to the data sets and for the semi-finals. The general room was split into multiple rooms, one per mentor, during mentoring sessions. Mentors entered the rooms and the participants scheduled meetings with them during those time slots or outside the dedicated time slots.

Slack was used to broadcast and to privately send messages and share information during the challenge. It was chosen because the platform allows knowledge and information management and enables members of the group to join conversations and access previous exchanges. Different channels were available: Common to all participants and mentors were a general channel for communicating organizational information, a help-desk channel for technical questions, and a channel for sharing questions regarding privacy issues. Private communication channels were opened for each team and were also used to communicate directly with mentors. A general public channel for communication with mentors was available. A private channel was created for mentors so they could conduct internal conversations, and a private channel was created for the organizing committee. The teams also used their own forms of communication (mostly Zoom, WhatsApp, and email) in addition to Slack.

## 2.9 Participant and mentor surveys

The validation of the methods was done via an anonymous questionnaire distributed to all participants and mentors one week after the event in an email sent to participants) and to mentors with a link to Google Forms. The Ethics Committee of the University of Haifa approved the study. The participant and mentor surveys were extensions of surveys used in Braune et al. [5] (Appendices C and D). The extended questionnaires gather the participants' and mentors' experience and perceived value. The extended participant survey included 38 questions.

The mentors' feedback in Braune et al.'s questionnaire focused on the difference in their experience as mentors in the virtual event vs. face-to-face mentoring. Questions were added to gather the mentors' feedback on things that worked well and those that should be improved. The additional questions regarded the organization of the Datathon and the specific aims of the Datathon, including community building, and the mentor's interests in working with the teams further. The extended mentor questionnaire includes 15 questions.

# 3. Results

The results of the questionnaires and the analysis of the recorded presentations and Slack channel usage are presented in four sections. Section 3.1 provides descriptive data related to the participants' background. Three other sections follow the Datathon's three objectives: community formation (Section 3.2), production of useful data science projects (Section 3.3), and formulation of a process for future data-based collaboration between the government and researchers from all sectors of society (Section 4.4). The latter is based on our analysis of points of strength and concern that are reviewed in the Discussion section.

## 3.1 Participants' background

Of 280 people registered for the Datathon, 78 participants passed the selection process and participated through the entire event. The participants were organized into 18 teams (6 per challenge). Participants were from different sectors: industry, mostly high-tech (40), academia (22), public sector (10), and health care (6). About 75% of the participants registered as part of a team. Of the 18 teams that started the Datathon, 7 teams were original teams, and the rest of the teams were assembled by the MoH based on individual registrations or registrations of very small teams. One team dropped out during the first day of the challenge and one team decided to split into two. Thus, there were 6 teams in the first challenge, 5 in the second challenge and 7 in the third challenge. Of the 280 registrants, 79 were female 9 (28%). Of the 78 participants, 25 were female (32%).

Online surveys were completed by 18 of the 78 participants (23%) and 12 of the 36 mentors (33.3%). The participants who answered the survey are representative of the body of participants in terms of the sectors to which they belong and in terms of having registered as teams vs. individuals. Most of them heard about the event via their social networks. Around 60% of the participants had attended hackathons in the past, and 30% of them had attended online hackathons before. Close to half (47%) of the respondents indicated that they have previously worked with the publicly available COVID-19 data (a subset of the full data sets provided during the Datathon).

## 3.2 Community formation and trust

### Online experience

Most of the participants were satisfied or highly satisfied from the online Datathon experience, including the participant selection process, the team formation process, the support provided by the organizing team and by the mentors, the adequacy of judging, and the data science and presentation tools supplied (Table 3). More participants expressed satisfaction from the data supplied than dissatisfaction. Half of the participants expressed concerns regarding the adequacy of the virtual environment (see specific points of concern and opportunities for improvement regarding the data sets and the virtual environment in sections 4.3 and 4.4).

Table 3. Participants' responses to survey questions regarding satisfaction of the Datathon. A bar chart on the right visualizes the responses.

| Question: How satisfied were you from the | 1 – highly unsatisfied | 2 – not satisfied | 3 – neutral | 4 – satisfied | 5 – highly satisfied | |
|---|---|---|---|---|---|---|
| Participant selection process | 0 | 1 | 6 | 3 | 6 | 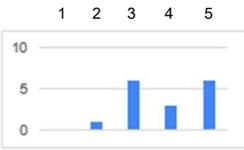 |
| Team formation process | 0 | 1 | 6 | 1 | 7 | 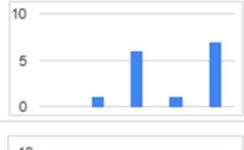 |
| Needs addressed by the organizers | 1 | 2 | 3 | 5 | 6 | 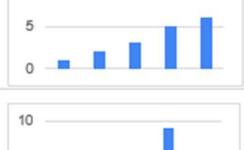 |
| Mentoring sessions | 0 | 1 | 1 | 9 | 5 | 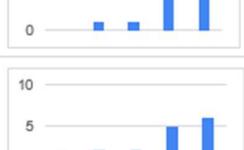 |
| Judging | 1 | 2 | 2 | 5 | 6 | 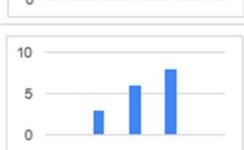 |
| Data science tools supplied | 1 | 3 | 0 | 6 | 6 | 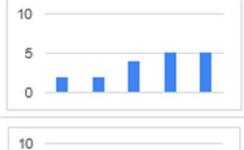 |
| Presentation tools supplied | 2 | 2 | 4 | 5 | 5 | 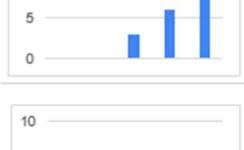 |
| Slack | 0 | 0 | 3 | 6 | 8 | 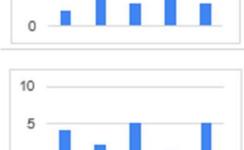 |
| Data supplied | 2 | 4 | 3 | 5 | 3 | 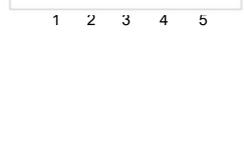 |
| Adequacy of the virtual environment | 4 | 2 | 5 | 1 | 5 | 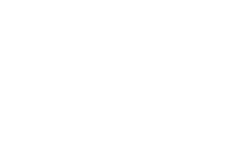 |

Participants missed the physical presence of the team members and mentors (and found that the status meetings that opened and closed each day were unable to replicate the physical experience). Mentors noted that while the online format offers greater availability, the physical environment, as one mentor noted, allows for "getting a glimpse of more projects". Four of 11 mentors reported that this difference negatively impacted the quality of the mentoring they were able to provide, because of the limited opportunity to be proactive.

Having said that, half of the mentors who answered the survey expressed high (4) or very high (2) rating of their mentoring experience. The majority of the mentors (6) reported that new insights were gathered. Two mentors reported learning novel facts about the COVID-19 data itself; one mentor highlighted the importance of integrating ethical considerations into the pre-event workshop (see Table 1).

### Collaboration during the Datathon

Most of the participants expressed high or very high satisfaction from the interaction with the mentors and from the judging (see Table 3). Most mentors (10 of 12) reported that it was easy to schedule meetings with the teams and that the teams' responses to their suggestions were satisfactory (5) or highly satisfactory (1).

### Use of Slack for collaboration

According to the survey results, most of the participants were satisfied or highly satisfied from the use of Slack for communication among team members and with the mentors and organizers (Table 3). Appendix E presents data on the usage of the Slack channels. Overall, Slack was routinely used by all participants, the mentors, and the organizers, but its main use was for announcements on the general Datathon channel (249 messages) and on the help-desk channel (333 messages).

Most of the mentors used the Slack channels to communicate on the general Datathon channel and to communicate with the organizers. Instead of using the help-desk channel, mentors relied on direct communication with the organizers. Most mentors did not document their advice on Slack; hence mentors that advised the teams could learn about the advice given by other mentors only through comments made by the teams themselves.

### Collaboration after the Datathon: community building and trust

Most of the participants and mentors who answered the survey expressed neutral opinions regarding their interest in sustaining the community that formed during the Datathon.

76% of the participants stated that without the Datathon, they would not have initiated their own COVID-19-related project and participants expressed higher confidence in starting their own venture in health care or digital health research after the Datathon experience than prior to it.

More than half of the participants who answered the questionnaires expressed interest in pursuing their project further. Among the winning teams, the team that won first place asked to continue working on their project to improve their model and generate deeper insights for potential publication. Their employer allocated 20% of their time to continue working on the project. Subsequently, the team wrote a paper about their project together with a researcher from the MoH [14], which helped teams within the government improve the vaccination policies by redefining the immunized population and could potentially help other governments implement similar policies. The Datathon second place team, together with leaders from the MoH, are still engaged in dialogue regarding how to use their insights to improve future policies The team who won third place did not have time to engage with the project further.

### 3.3 Projects - helping solve the challenges

Appendix F presents information regarding the six finalists: the teams and their projects, including team composition; prior experience with COVID-19 data; #mentors sought by the team; project title; data science methods and tools used (and the judges' ratings for this aspect); main insights; and the judges' ratings of the potential impact of the project, degree of innovation, ease of implementation considering regulatory aspects, and ease of working further with the team itself. Table 4 provides more details on the top three projects, the first two of which concerned the young population challenge and the third concerned compliance to COVID-19 tests.

Table 4. Details on the three top projects

| Project 2.3. Policies for children in a vaccinated reality |
|---|
| The team envisioned a decision-making tool for individuals, school principals, and city mayors; given information on the COVID-19-related morbidity (%positive, %deaths), the demographics (population density, median age, GDP, socioeconomic level, reproduction rate), vaccination rate (#vaccinated 1st dose, #vaccinated 2nd dose), and a selected policy, the model would predict the projected morbidity. Thus, citizens living in a certain statistical region could test different behaviors (e.g., how many people they would like to meet) and view the implications, and decision-makers could get a recommendation on the best policy. The policies may be related to schools, labor, economic aid, hygiene, flights, and gatherings. The team developed the model using AdaBoost from the Oxford data set of 250 countries and were able to make accurate predictions using the Data.gov data set for the first 20 days of the pandemic. |
| Project 2.1 How can the isolation period in the young population be shortened? |
| Some exposures to the virus do not result in active infections. If an infected individual exposed a group to an infection, then the probability that the infection is inert grows as the number of individuals in the group test negative. Given a level of false negative risk that we are willing to accept, we can easily calculate the size of the subgroup that needs to be tested and the day on which testing should be done following exposure. If the group goes to isolation and some members of the subgroup are tested n days after exposure and they are all negative, then the entire group can be released from isolation, otherwise they are retested n days later. The exposure announcement often arrives 3-4 days after infection, and by testing 5 children and getting results while the kids are still at daycare, they might already be isolation-free by the end of the day! |
| Project 3.4 Targeted testing and incentive allocation |
| The team had previously developed methods for uplift modeling [11]: using incentives to increase the probability of an intended behavior, in this case, getting tested for COVID-19. |

> Incentives target the persons that are most likely to generate value from the incentive: those who are (a) more likely to be positive when tested and (b) would not go to get tested without incentives but are likely to respond to personalized incentives. They therefore target persons who have had many connections in epidemiological investigation, which the team plans to extend to social centrality. This value is weighted with the probability of testing positive on the first test. Features that predict this include the timing since the beginning of the epidemic, the age of the person, and the locality. They propose to divide a target population into groups, test different incentives, collect data, and build and update a model.

# 4. Discussion

## 4.1 Main findings

In this study, we have shown for the first time, that for an important cause it is possible to successfully engage a diverse community in a datathon that transparently shares up-to-date government data. Consequently, government policies were influenced by insights developed by the first place team [14]. Additionally, a process for future research collaborations was established, spanning all sectors: national and local governance, academia, the high-tech industry, and health care organizations.

A highly significant finding is that 12 of 16 participants who responded to the survey indicated that the event had highly increased their trust in the government. This matter is of considerable significance for two interrelated reasons: public confidence is an essential element for maintaining resilience in the face of an emergency [15-18], and public confidence in many jurisdictions (including Israel) has suffered for various reasons (the examination of which requires a different study) [19,20]. The matter of access to novel data was also reflected by several of the participants' comments as one of the key features that made the Datathon worth their while.

Analysis of the Datathon's results and the surveys revealed points of strength as well as limitations and allowed us to learn from this experience and create a blueprint for future such datathons.

## 4.2 Points of strength

### Diversity of participants

The open call and the selection process, which was sensitive to diversity, resulted in engaging participants from a broad range of disciplines, from academia, industry and the public sector, from different religion backgrounds. About a third (32%) of participants were women (28% of registrants were women).

### Organization and personnel

The MoH's process that assembled mixed teams, possessing complementary expertise, was successful. Half of the finalist teams, including the winning team, were mixed.

The vast majority of the respondents indicated that their needs and questions were met by the organizing team.

### Data science tools provided

Most of the participants were satisfied or highly satisfied with the tools provided. Participants suggested adding some specific R and Python packages and other tools that they use at work, including Amazon and Azure tools, Matlab, and SAS.

## 4.3 Points of concern

Future organizing teams will have to consider their preferred optimization when facing the following points of concern.

### The participant selection process

As the transparency regarding the selection process increases, accountability and potential for trust increase. Yet, embracing transparency to the point of providing explicit ratings for each of the applicants on each of the criteria listed risks dissuading those who received lower ratings from further engagement and raises the specter of an adversarial process of contestation. The organizing team opted for "soft" grades of three levels: accepted, waitlisted, or rejection.

### The selection of challenges

By focusing on three specific challenges established in a process involving multiple stake-holders, other problems were not available for research, and some participants have suggested that a longer list of more concrete problems would have been better. An interesting idea mentioned in the surveys is to consider a "wild card" challenge where teams come up with their own challenge and try to solve it. For such a challenge, an all-inclusive data set could be constructed.

### Data and models standardization and sharing

A point of strength of the government-shared data set was that it was standardized based on the Israeli Bureau of Statistics codes, which could make it joinable with other tables. However, due to privacy concerns, steps have been taken to prevent such joins, honoring privacy while sacrificing some of the data's potential. Data belonging to medical groups is richer than the government's data. For example, it includes patient diagnoses, which can allow analysis of risk factors. However, such has not been shared with the public.

To create an integrated database, standard definitions should be provided for concepts such as seriously ill patients, for policies of lockdown and school opening, etc. This is not easy to do even within a single country, because these definitions have changed over time. In our particular Datathon, such temporal definitions were mediated by supplying the worst state of a hospitalized patient over the entire duration of a hospital stay. Patient data standards that define semantics, such as SNOMED-CT codes, Observational Medical Outcomes Partnership (OMOP) Common Data Model, and communication protocols (e.g., Health Level Seven's Fast Healthcare Interoperability Resources (FHIR)) could help in solving technical challenges. Yet the government-to-government process of data integration and sharing entails political and regulatory challenges as well and is a topic for future research.

Some participants have expressed a desire to continue to work on the data and further advance their proposed insights. Note that, to reduce the chance of a privacy breach, all work was conducted on the MoH's virtual rooms. Participants were not able to download the data, code, or models to their computers without review of the organizing team. Given the ability of a complex deep learning module (DNN or CNN) to encode private information, we limited their export even for teams who created them, which made the use of these modules even less favorable. This raises the issue of finding ways to continuously share the data.

### The timeframe and the infrastructure

This Datathon lasted for 48 hours, which offered the ability to present a basic concept and test it on the available data, but not much more than that. Related in part, is the research environment. While conventional hackathons are premised on having access to any and all tools and available data, this Datathon was held in a controlled environment, striking a balance between usefulness on the one hand and data security and privacy on the other. This balance resulted in slower response time and unavailability of some tools. Consequently, some participants wished for more time. It seems that an additional 12 hours would have made a favorable difference for many.

### Prizes and recognition

The incentives to participate in a datathon are multiple. Some look for important puzzles to solve, especially if the data is unavailable elsewhere. Others seek networking. Yet others seek to develop prototype solutions that could be turned into products and sold. Given the very tight budget this Datathon operated under, the organizing team decided that no monetary prizes would be awarded. However, there is a selection bias because people who sought monetary prizes likely did not join the Datathon to begin with.

## 4.4 Lessons learned: how to improve the process

### IT and tools

Future datathons should consider including a slot every 24 hours where the environment can be updated with software requested by the participants and approved by the organizers' IT team. Needless to say, 24/7 IT support is required throughout the event itself.

### Data

Based on our experience, we recommend adding a pilot study serving as a quality assurance step, two weeks prior to the event (Table 1). In this pilot, a team which would not compete in the Datathon would test the research environment and the data sets before the actual participants would be invited to test the data. Such strategies are important in order to avoid the participants' impression that some challenges could not be solved in a meaningful way with the data provided, or that the government was simply trying to show how difficult the policy challenges are, without providing all the data that it actually has, arguing that such sharing would compromise citizens' privacy.

### Improving transparency

We suggest that the announcement of the Datathon should disclose the limitations on the number of participants, and the lack of budget for prizes and food. This could have lowered some resentment both of registrants who were not selected to participate and of participants.

### Media attention and publicity

Given the importance and novelty of the event (perhaps the first government-led datathon related to COVID-19 data), it is important to generate a clear plan for engaging the media—traditional and social, before, during and after the event. Ideas include interviews with participants, mentors, judges, and the organizing teams, follow-ups after some time, and reports by the winning teams on actual implementation of their insights.

### Participant communication process and interaction with the mentors and organizers

Mentors reported difficulty in setting up mentoring sessions via Zoom or Slack, difficulty in realizing which participants belonged to which team, lack of documentation of the advices given during mentoring sessions, and being idle. Most of these concerns could potentially be at least partially met by using the Mentornity platform [20] suggested by the organizers of a COVID-19 hackathon [5]. This tool shows mentors' profiles, which can be searched by areas of expertise, and facilitates video call sessions for each team. Braune el al. [5] required each team to book, via Mentornity, at least one mentoring session. The number of mentoring sessions held by every mentor and team can be easily tracked by this platform, which also provides feedback forms to summarize the mentoring sessions. This may allow mentors to learn from one another and may allow the judges to reflect on the suggestions provided by mentors.

The daily status updates were important for ensuring the flow of the event and for communicating all-challenges announcements and updates. Furthermore, they provided an opportunity for the teams to present themselves. In the future, such status updates could also include some suggestions from mentors that may be helpful for a given challenge (or for all challenges).

*Judging—real-time and ex post communication*

Participants of the semi-finals could have received greater input from judges about the strength and weaknesses of their solutions (beyond learning whether they passed to the next stage or not). Some have expressed concerns that the judges overemphasized the "commerciality" or "time-to-market" aspects of their solutions, but given the lack of ex post debriefing with judges or pointers for further ways to pursue their ideas, participants were left unsure.

*Improve community building*

Participants noted that a dedicated social group for the event could be formed, which may then carry forward to future events. Mentors observed the importance of informing the audience of future planned events and steps, subsequent to the Datathon.

*Integrate assessment survey into the closing event*

This time window is preferred because participants are waiting to learn who won the competition and thus are still behind a veil of ignorance of sorts, while still fully engaged.

## 4.5 Process for future collaborations between the government and the research community

We recommend that future public-private datathons would follow the process detailed in Table 1 that builds upon the original process undertaken in this project, modified by lessons learned. Attention should be paid to expanding datasets released for analysis, conditional upon deployment of deidentification processes in a controlled environment. Investing in the technical infrastructure (in terms of bandwidth and processing power) and in providing access to a wide range of data science tools are also important factors for successful future collaboration. Finally, carefully designed project-management processes, premised on constructive communication among the various circles of stakeholders and fit with the regulatory and ethical environment, are key.

## 4.6 Conclusion

The Datathon, which solicited participation of a rich body of researchers from relevant sectors, proved to be a promising start at gathering innovative insights to policy-related challenges. The MoH is currently exploring ways to utilize these solutions. Moreover, and at least equally importantly, the participants appreciated the access provided by the government to a relatively large data set, which expressed the implemented commitment of the government to balanced transparency—itself a key component of sound policy. These elements of transparency and the spirit of a joint venture were reflected in an increase in public confidence pursuant to the Datathon, as reported by the participants.

The process analyzed in this paper is replicable by other governments, and in contexts other than the COVID-19 challenge. In those contexts, thoughts should also be given to government-to-government (G2G) platforms, since expanding data sets may enrich the usefulness of the tool development (and also generate more general social goods related to G2G cooperation). This paper is therefore an invitation to replicate and refine the process discussed here and further experiment in constructive tools for a productive data-sharing policy.

*Acknowledgement*

Part of this work was made possible by resources made available by the Data Science Research Center at the University of Haifa (grant number 100008584 from the Council for Higher Education) and by the Center for Cyber, Law, and Policy at the University of Haifa in conjunction with the Israel National Cyber Directorate in the Prime Minister's Office.